# Structural, Elastic and Optoelectronic Properties of inorganic cubic $FrBX_3$ (B = Ge, Sn; X = Cl, Br, I) perovskite: the Density Functional Theory Approach

Nazmul Hasan[a], Md Arifuzzaman[b*] ,and Alamgir Kabir [c*]

[a]Department of Electrical and Computer Engineering, North South University, Dhaka 1229, Bangladesh.
[b]Department of Mathematics and Physics, North South University, Dhaka 1229, Bangladesh
[c]Department of Physics, Dhaka University, Dhaka 1000, Bangladesh.

*Corresponding author: alamgir.kabir@du.ac.bd (A. Kabir)
   md.arifuzzaman01@northsouth.edu (M. Arifuzzaman)

**Abstract:**
Inorganic metal-halide cubic perovskite semiconductors have become more popular in industrial applications of photovoltaic and optoelectronic devices. Among various perovskites, lead-free materials are currently most explored due to their non-toxic effect on the environment. In this study, the structural, electronic, optical, and mechanical properties of lead-free cubic perovskite materials $FrBX_3$ (B=Ge, Sn; X= Cl, Br, I) are investigated through the first-principles density-functional theory (DFT) calculations. These materials are found to exhibit semiconducting behavior with direct bandgap energy and mechanical phase stability. The observed variation in the bandgap is explained based on the substitutions of cations and anions sitting over B and X-sites of the $FrBX_3$ compounds. The high absorption coefficient, a low reflectivity, and a high optical conductivity, make these materials suitable for the photovoltaic and other optoelectronic device applications. It is observed that the material containing Ge (germanium) in B-site has higher optical absorption and conductivity than Sn containing materials. A systematic analysis of the electronic, optical, and mechanical properties suggests that among all the perovskite material, $FrGeI_3$ would be the potential candidate for optoelectronic applications. The radioactive element Fr contained perovskite $FrGeI_3$ may have applications in nuclear medicine and diagnosis such as X-ray imaging technology.

***Keywords:*** Inorganic Perovskite, Metal-halide, Semiconductor, Optoelectronics.

1. **Introduction**

Metal-halide cubic perovskites are a class of semiconductors represented by the chemical formula $ABX_3$, (where A is a cation, B is a divalent metal ion, and X is a halogen anion). Recently, they have gained the scientific community's enormous interest because of their high-efficiency solar cell applications compared to silicon-based technology.[1]–[6] Moreover, they are found to unveil remarkable optoelectronic properties, such as tunable direct bandgap, long charge diffusion length, spanking charge carrier mobility, low carrier recombination rate, small excitation binding energy, high optical absorption, and high dielectric constant, which make them more competent in various device applications including light-emitting diodes (LEDs), photovoltaic, and solar to fuel energy conversion systems and solar cells.[7]–[10] It is reported that the hybrid perovskite has been demonstrated a remarkable improvement over the photoconductor-based X-ray detector [11]–[13], spectroscopy[14], acoustic wave signal processing, and image storage devices [15]. A profound understanding of the characteristics of this class of exciting materials is required to forecast their specific device applications. As a result, it is indispensable to explore the structural, mechanical and optoelectronic properties and overall features of different perovskites.

Lead halide perovskites with the cubic structure have sparked immense attention because of their potential use in solid-state solar-cell systems with high power conversion competencies and relatively inexpensive and straightforward processing.[16]–[20] Although most of the naked optoelectronic features of these materials have already proven them as potential candidates in optoelectronic technologies with the electronic, optical and maximum throughput, their toxicity issue bares them to be commercially successful. Moreover, lead-based metal halide perovskite materials dissolve to $PbI_2$ in the presence of oxygen in the atmosphere at ambient conditions, which is environmentally detrimental.[21]–[23] Thus, one of the major drawbacks of lead halide cubic perovskites include the instability of devices and J-V hysteresis.[24] One major challenge to commercialize the perovskite solar cell is to make lead-free non-toxic halide perovskite.

There are studies to demonstrate the methods in forming thin films with Cesium (Cs) at A-site, projecting the significant enhancement in efficiency for the fully inorganic lead halide perovskites by utilizing a mixed halide composition in $CsPbI_3$ material. [4] The structural, optoelectronic and vibrational properties of $CsPbCl_3$ are investigated by *Padmavathy, R. et al.*[6] and found that the materials are wide-band-gap semiconductors as well as dynamically unstable. Another study performed by *Padmavathy, R. et al.* investigated the properties for lead-free $CsSnI_{y-y}Cl_y$ perovskites. They demonstrated the band gap tunability by pressure variation along with material's phase transition which was projected for UV light applications. [7] *Roknuzzaman et al.* performed a thorough DFT investigation of a series of metal halide perovskite materials ($CsBX_3$ (B = Ge, Sn, and X = Cl, Br, I) and $CsPbX_3$ (X = Cl, Br, I)) employing the first-principle study.[25] They investigated the structural, electro-optical, and elastic properties of Ge-based $CsGeI_3$ materials and recommended them as the best class of perovskites after lead-free metal-halide perovskites for solar cell and optoelectronic applications. *Jellicoe et al. conducted the experimental* investigations on the absorption profile and photoluminescence of $CsSnX_3$ (X = I, Br, Cl) and performed the theoretical calculations on their optical characteristics.[26] Recent advancements of lead-free metal halide perovskites for photovoltaic applications as well as modulation in energy band gaps along with optoelectronic and elastic properties following different mechanisms (i.e., meal doping in B-site, hetero-valent substitution, pressure induce method) have been reported in the literature.[27]–[36]

However, no investigations are found yet in the literature on the lead-free perovskite compounds with considering Fr in the A-site for such $ABX_3$ series where (B= Ge, Sn; X=Cl, Br, I) due to its low natural abundance. $FrBX_3$ (B= Ge, Sn; X=Cl, Br, I) perovskite may be a gateway to understand the behavior of perovskite containing heavy radioactive element in A-site. Hence, herein we perform the first-principle calculations to study the structural, elastic, optical and ecteronic properties of inorganic lead-free metal-halide perovskites $FrBX_3$ through the Density Functional Theory (DFT) approach.

## 2. Computational Methodology

To uncover the structural and optoelectronic properties of FrBX$_3$(B = Ge, Sn; X = Cl, Br, I), we perform the Density Functional Theory (DFT) simulations based on the plane-wave pseudopotential using the Cambridge Serial Total Energy Package (CASTEP) code of Materials Studio-7.0 and the structural relaxation were re-confirmed by the Vienna *ab-initio* Simulation Package (VASP).[37]–[42] The constructed unit cells of FrBX$_3$ in the cubic form are shown in Figure 1, based on which further investigations in this study are carried on. The geometry optimizations are performed by employing the Generalized Gradient Approximation (GGA) for exchange-correlation interactions into the Perdew-Burke-Ernzerhof (PBE)[43], [44] functional. The electron-ion interactions have been studied using the Vanderbilt's Ultrasoft pseudopotential (USP) [45] with the Koelling-Harmon relativistic treatment. The Broyden-Fletcher-Goldfarb-Shanno (BFGS) algorithm [46] is used to secure the optimized crystal phase to cut down the total electronic energy, internal forces, and stresses. In these calculations, the wave function is prolonged up to 450 eV as the cutoff energy for the plane wave function. A Monkhorst-Pack's K-point mesh of 10×10×10 K-points sampling in the Brillouin zone is used to secure the better convergence. [42] Elastic stiffness constants (C$_{ij}$) were calculated by assuming the finite strain theory within the CASTEP code [40], [47]. The Voigt-Reuss-Hill (VRH) averaging scheme [48] along with the relevant equations [28], [49] are used to calculate the polycrystalline mechanical parameters by setting the maximum strain amplitude to 0.003. The optical properties are calculated and analyzed excepting the scissor operator regarding an obvious depiction using the CASTEP based DFT Kohn-Sham orbitals, and supported formulae are available in the literature.[50] In the present study, the optimization threshold for the geometry of the unit cells and atomic relaxation were set up in the CASTEP as follows:

- total energy 5×10$^{-6}$ eV/atom;
- maximum force 0.01 eV/Å;
- maximum stress 0.02 GPa;
- maximum displacements 5×10$^{-4}$ Å.

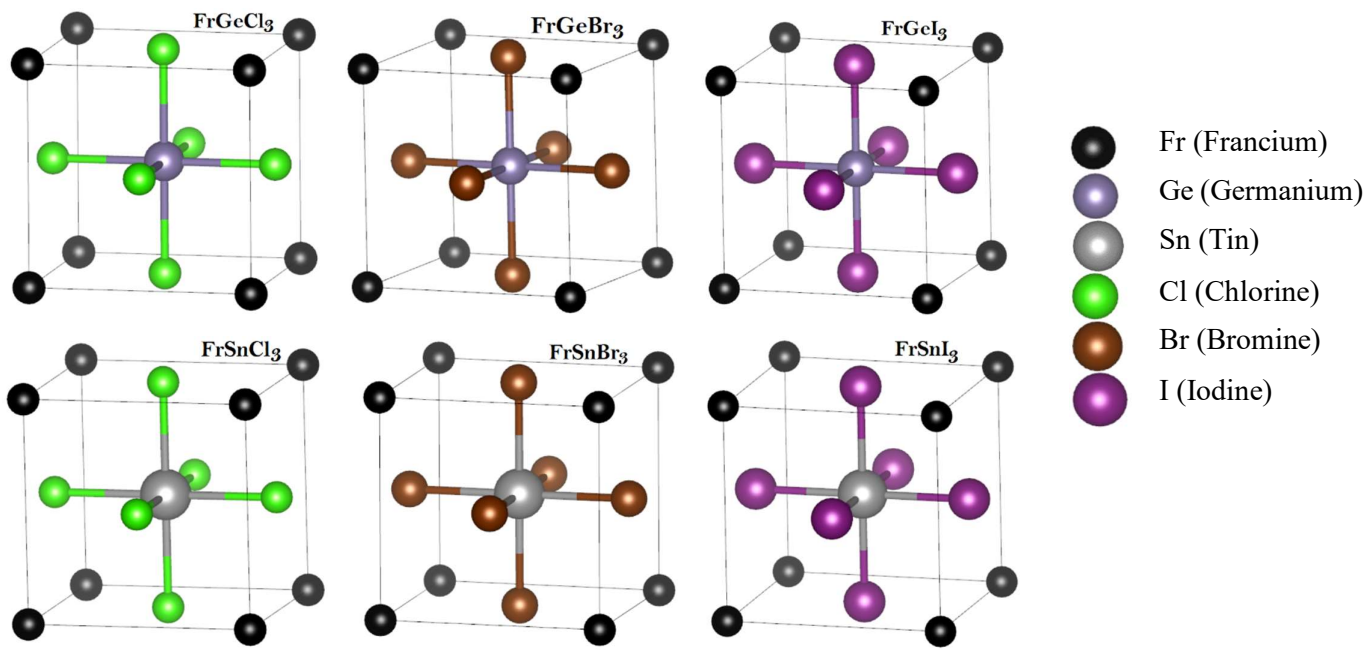

**Figure 1.** The considered cubic metal halide perovskites' unit cells of FrBX$_3$ (B = Ge, Sn; X = Cl, Br, I) followed by ABX$_3$ structure. The structures are optimized within the DFT approximation.

## 3. Results and discussion

### *3.1 Structural properties*

The investigated materials FrBX$_3$(B = Ge, Sn; X = Cl, Br, I) belong to the cubic crystal structure under the space group of *Pm$\bar{3}$m* (no. 221). These crystal structures have been modeled with the program of 3D Visualization for Electronic and Structural Analysis (VESTA). The unit cell of each material comprises five atoms, as depicted in Figure 1, where the Fr atoms occupy the corner positions of the cubes at 1-a Wyckoff coordinates of (0, 0, 0). The B-site cations are located at the 1-b Wyckoff site (0.5, 0.5, 0.5) in the crystal's body-centered positions, whereas halogen (X) atoms occupy the face-centered positions at the 3-c Wyckoff site (0, 0.5, 0.5).

The estimated lattice parameter and volumes of the optimized cells are listed in Table 1. The lattice parameter and unit cell volume are reduced when halogen atom (X) is replaced with another halogen atom of smaller ionic radius, whereas replacing germanium (Ge) by tin (Sn) in the B-site exhibits the opposite trend, as appeared in Table I. With the change of halides size, a clean periodic pattern is observed in the variation of lattice parameters due to the 'octahedral effect' in the cell volume, which upholds the universal trend of variation in atomic size and charge of the most perovskite materials.[24], [51], [52] Same periodical manners are also exhibited in the case of interatomic bond lengths of the materials as appeared in Table I. The substitution of one type of atom for another in the compound affects its structural characteristics and the material's performances, such as electrical, optical, and mechanical capabilities.

*Table I: The calculated unit cell and structural parameters of FrBX$_3$(B=Ge, Sn; X=Br, Cl, I).*

| | FrGeCl$_3$ | FrGeBr$_3$ | FrGeI$_3$ | FrSnCl$_3$ | FrSnBr$_3$ | FrSnI$_3$ |
|---|---|---|---|---|---|---|
| a(Å) | 5.37 | 5.63 | 6.01 | 5.64 | 5.90 | 6.27 |
| V(Å$^3$) | 154.72 | 178.42 | 217.45 | 179.83 | 205.82 | 246.30 |
| Inter atomic bond length (Å) | Fr-Ge: 4.65<br>Cl-Ge: 2.68<br>Fr-Fr: 5.37<br>Cl-Cl: 3.79 | Fr-Ge: 4.87<br>Br-Ge: 2.81<br>Fr-Fr: 5.63<br>Br-Br: 3.98 | Fr-Ge: 5.21<br>I-Ge: 3.01<br>Fr-Fr: 6.01<br>I-I: 4.25 | Fr-Sn: 4.88<br>Cl-Sn: 2.82<br>Fr-Fr: 5.64<br>Cl-Cl: 3.99 | Fr-Sn: 5.11<br>Br-Sn: 2.95<br>Fr-Fr: 5.90<br>Br-Br: 4.17 | Fr-Sn: 5.43<br>I-Sn: 3.13<br>Fr-Fr: 6.27<br>I-I: 4.43 |
| Inter atomic bond angles (°) | Fr-Ge-Cl: 54.74<br>Cl-Ge-Cl: 90 | Fr-Ge-Br: 54.74<br>Br-Ge-Br: 90 | Fr-Ge-I: 54.74<br>I-Ge-I: 90 | Fr-Sn-Cl: 54.74<br>Cl-Sn-Cl: 90 | Fr-Sn-Br: 54.74<br>Br-Sn-Br: 90 | Fr-Sn-I: 54.74<br>I-Sn-I: 90 |

As different cations can be incorporated into the ABX$_3$ cubic perovskite framework resulting in the development of diverse materials with identical properties, so the tolerance factor has a significance to discovery of alternative lead-halide perovskites. Concerning with perovskite materials' efficiency with solar conversion, instability in phase in ambient environment and rapid crystallization during the fabrication process are highly effect. Following the no-rattling principle for the perovskites, the perovskite forming factor by geometric conditions empowers us to anticipate perovskites with an 80% fidelity. The tolerance factors for the investigated perovskite materials are calculated using the Goldschmidt's tolerance factor following the equation: $T = (R_A+R_X)/\sqrt{2}(R_B+R_X)$, where $R_A$, $R_B$, $R_X$ stands for the ionic radii for A, B, and X site ions respectively. It is observed from the Table II that the tolerance factor is varied in range of 0.84 to 1.00 in which Ge based materials' values exhibit close to 1 to strongly recommend that the formation of an ideal and tightly packed stable cubic perovskite structure having size of cation A larger than that of B [53], [54].

*Table II: Comparison of evaluated tolerance factors for the FrBX₃ (B= Ge, Sn'= Cl, Br, I) perovskites with the necessary parameters.*

| Element | T | IFF | $R_A$ | $R_B$ | $R_X$ |
|---|---|---|---|---|---|
| FrGeCl3 | 1.00 | 0.65 | | $Ge^{2+}$ | $Cl^-$ |
| FrGeBr3 | 0.99 | 0.68 | | 0.73 Å | 1.81 Å |
| FrGeI3 | 0.96 | 0.74 | $Fr^+$ | | $Br^-$ |
| FrSnCl3 | 0.85 | 0.58 | 1.80 Å | $Sn^{2+}$ | 1.96 Å |
| FrSnBr3 | 0.85 | 0.61 | | 1.10 Å | $I^-$ |
| FrSnI3 | 0.84 | 0.66 | | | 2.20 Å |

### 3.2 Electronic properties

The electronic properties are primarily analyzed considering the high symmetry directions of the Brillouin zone. The electronic band structures and density of states (DOS) of the pristine $FrBX_3$ (B=Ge, Sn; X=Br, Cl, I) compounds have been calculated through the PBE functional assuming the GGA approximation. The band structure of the $FrBX_3$ is shown in Figure 2. The energy band is considered from -5 eV to +5 eV in this study, and zero-point energy is taken as Fermi energy. It is clear from Figure 2 that all six structures exhibit the direct bandgap. The value of the bandgaps is reported in Table-III. The range of the bandgap energy indicates that the structures are semiconductor in nature. The valance band and bottom of the conduction band of all structures are located in the high symmetry R points in the Brillouin zone, which confirm the direct bandgap semiconducting nature of the $FrBX_3$ perovskite material. The retrieved bandgaps and band structures of the investigated materials suggest their novelty for photothermal, photovoltaic, and other optoelectronic applications. According to the semi-conductive theory, the nature of the materials can be confirmed precisely through their band structure patterns with reference to the Fermi level. Table III presents the calculated electronic band gap values for the compounds in this study, which are the characteristics of other perovskite materials containing $ABX_3$ structure as found in the literature. [7], [15], [20]-[24], [27]-[30] The present study is focused on observing the variation in bandgap due to the substitution of cations and anions over B- and X-sites of the $FrBX_3$ compounds.

Table III: Electronic Bandgaps (Eg) of the pristine FrBX$_3$ (B=Sn, Ge; X=Br, Cl, I).

| Samples | Electronic Bandgap, Eg (eV) |
|---|---|
| FrGeCl$_3$ | 1.14 eV |
| FrGeBr$_3$ | 0.81 eV |
| FrGeI$_3$ | 0.64 eV |
| FrSnCl$_3$ | 1.05 eV |
| FrSnBr$_3$ | 0.67 eV |
| FrSnI$_3$ | 0.42 eV |

The variation of energy bandgap with respect to the change of halogen in the X site are shown in Figure 2. The electronic band gap energy value decreases as we replaced Ge by Sn on B site for the same halogen atom, with a slight shift of conduction band towards the Fermi level in the band structures as observed in Figure 2. This decrease in the bandgap may be attributed to the increase of cell parameters for the FrSnX$_3$ compound as compared to the FrGeX$_3$ compound. As we change the halogen atom in X site for the FrGeX$_3$ element, a decrease in bandgap with larger halogen atoms is observed as in the Figure 3. The trend is similar for the FrSnX$_3$ compound. The interatomic distance will aggrandize as the lattice constant increases. As a result, the valence electrons' binding forces will weaken which will be lied in the valence band. However, bounding valence electrons require energy to move freely within the material and become conduction electrons and the energy gap is the minimum energy required to convert valence electrons to conduction electrons. As the interatomic distance increases, the valence electrons become less bound, requiring less energy to free them in the conduction band. However, the compositional dependence of the lattice constant for lattice mismatch at the interfaces of heterostructures, as well as the periodic potential field of the material, play an important role on the wave function of the valence electrons in a semiconductor crystal, which leads to variation in the energy band gap [55]. Furthermore, the energy gap is inversely proportional to the dielectric constant, which is inversely proportional to the inter-atomic distance. Consequently, increasing the lattice constant results in the decrease of the electronic band gap.

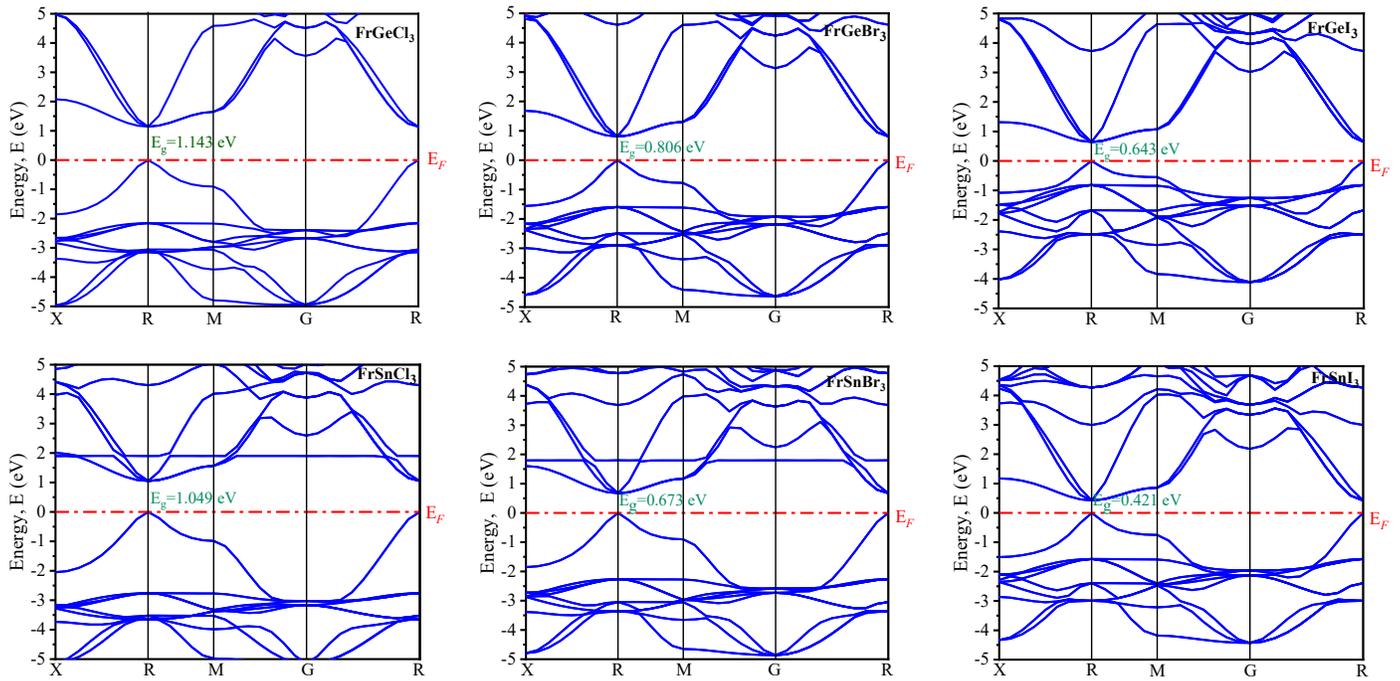

**Figure 2**. Electronic band structures of FrBX$_3$ (B = Ge, Sn; X = I, Br, Cl).

The tunability of band structures and observed bandgaps due to the change of atoms over B- and X-sites in metal halide cubic perovskite materials have made them prominent for various photovoltaic applications.

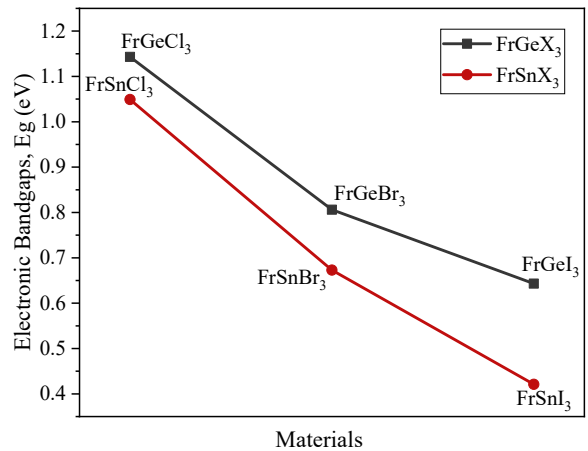

**Figure 3**. Variations of the electronic bandgap due to the effect of B- and X-site substitutions.

The calculated atom projected density of states are presented in Figure 4(A)-4(F). The total density of states (TDOS) and orbital projected density of states (PDOS) of each materials and constituted elements are available in *supplementary Figures (S2-S8)* for the pristine FrBX$_3$. From Figure 4, it is clear that the halogen atoms contribute more on the bottom of the conduction band, and the metal on the B-site (Ge or Sn) contributes more on top of the valance band. The Fermi level is situated adjacent to the top of the valance band. The orbital projected DOS (*supplementary Figure S3-S8*) revealed that at the Fermi level, the main contribution comes from the p-orbital of the halogen atom and the valence band maxima composed of the p-orbital of the Fr atom. The replacement of atoms in B site by Sn results in the slight increase of electronic states in the valence bands (VBs) below the ($E_F$) Fermi level. However, the change in halogen atoms leads to demonstrate opposite trends in electronic states. The investigated systems' conduction bands (CBs) result from halogens' and B-site atoms' p orbital, and the Fr atom contribute very less at the Fermi level. The significant contribution comes mainly from the Fr atom in the deep energy level between -8 eV and -6 eV. No hybridization in the generation of energy levels in the Fermi level is noticed for all configurations.

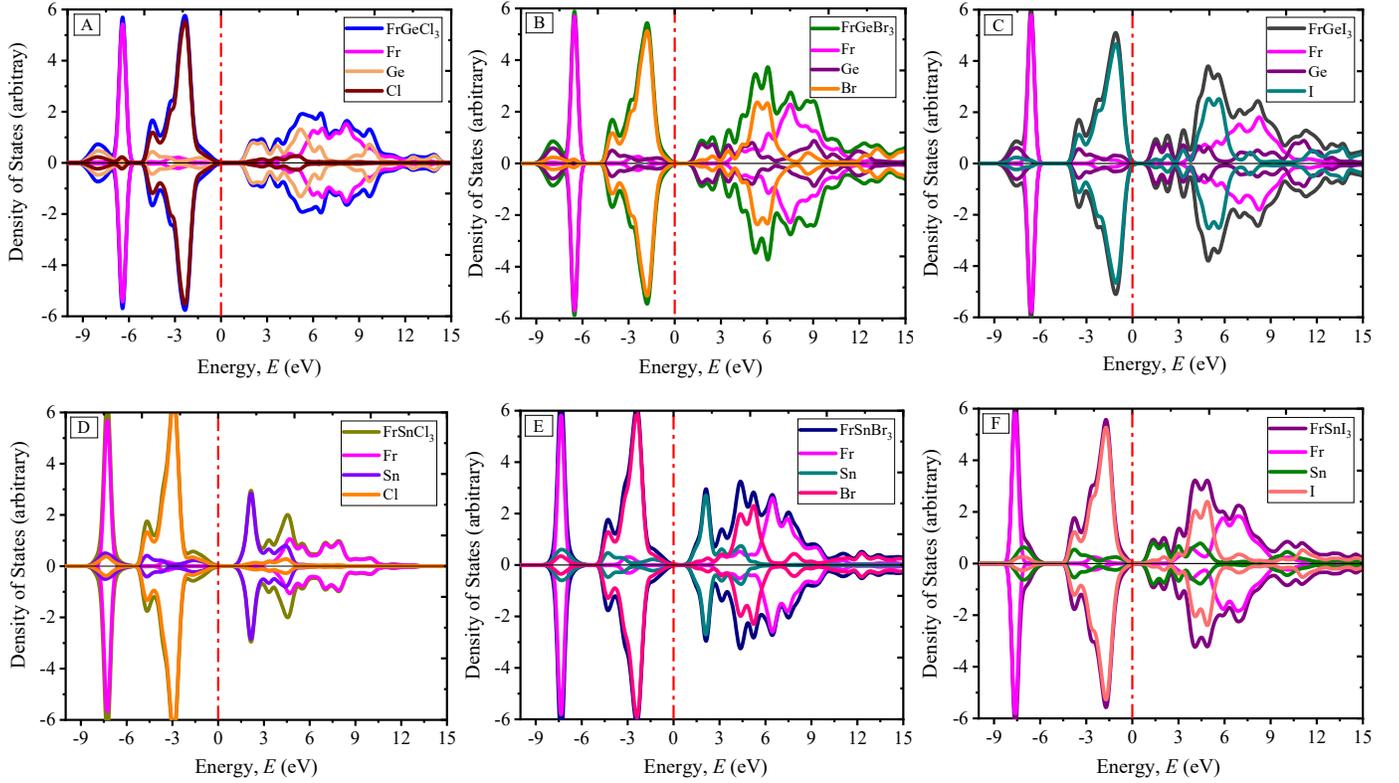

**Figure 4**. Atom projected electronic density of states (DOS) of perovskites $FrBX_3$ (B = Ge, Sn; X = I, Br, Cl).

### 3.3 Optical properties

A material's optical characteristics are the most fundamental parameters for its competence in optoelectronic and photovoltaic applications since they provide essential information about its interaction with light. Optical functions of a material are essential to the better understanding of the materials' electronic configuration and determine their suitability in photovoltaic applications. Therefore, the detailed optical properties of the considered cubic perovskite materials, such as absorption spectra with respect to the light energy and wavelength, reflectivity, refractive index, dielectric constants, and optical conductivity, are examined up to 30eV of photon energy.

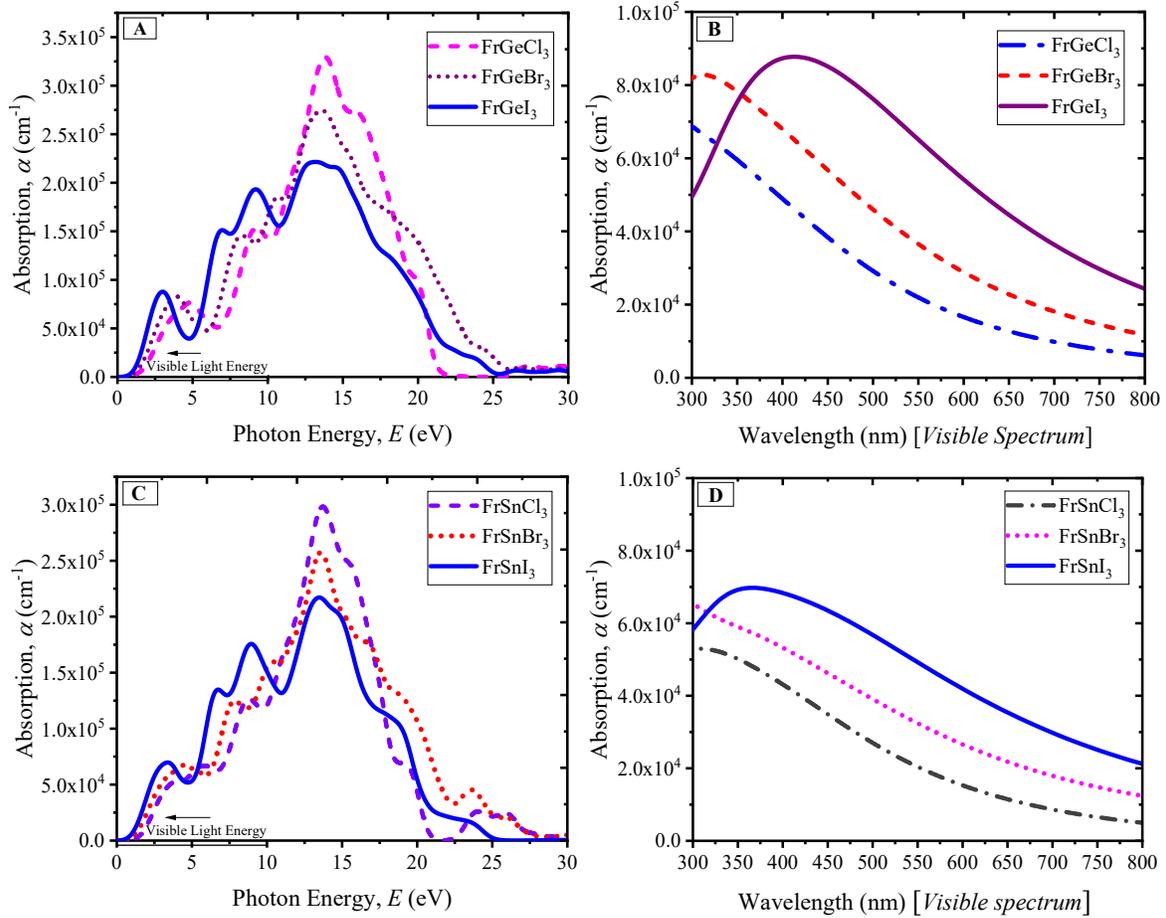

**Figure 5**. Absorption spectra of $FrBX_3$ (B = Ge, Sn; X = I, Br, Cl) as a function of wavelength and photon energy.

The optical absorption coefficient ($\alpha$) indicates how much light of particular photon energy or wavelength penetrates the substance before being absorbed by the material. That provides vital knowledge on the material's optimum solar-energy conversion efficiency for practical applications. In Figure 5, the evaluated photon energy-dependent (Figure 5(A) and 5(C)) and wavelength-dependent (Figure 5(B) and 5(D)) absorption profiles of the considered pure compounds are demonstrated. Generally, three significant peaks are observed for the compounds, as shown in Figure 5(A) for the $FrGeX_3$ compound, where a small peak at 6 eV photon energy is observed. The small peak at 6 eV photon energy becomes prominent for the case of the FrSnX$_3$ compound as shown in Figure 5(C). Changing in B site by Sn also results in the amplitude of the absorption spectra decreasing, keeping the peaks position at the same energy level. The highest peaks for the all-pristine compounds are found at ~14eV, whereas the second highest peak is found at ~9eV for FrGeX$_3$ compounds, and a slight change towards the low energy region (<10eV) is observed for FrSnX$_3$. However, the highest peaks are found for Cl-containing materials at the halogen site at ~14eV as observed from the absorption profiles in Figure 5(A) and 5(C). The first peak for all compounds is observed from 2.5eV to 5eV in the low-energy region. Changes in halide atoms affect the absorption magnitudes; I-containing materials (FrB[=Ge,Sn]I$_3$) exhibit more absorption at low energy regions and a wider absorption profile compared to Br- and Cl-containing compounds. The higher absorption in the low-energy region indicates the material to be a probable candidate for the solar cell application since the sun radiates more energy in the visible region. The absorption profiles with respect to the wavelength are exhibited in Figure 5(B) and 5(D) that are projected to further understand the light absorbance behavior of the particular FrBX$_3$ series compounds in the visible range of the light spectrum. According to Figure 5(B) and 5(D), I-containing materials demonstrate a comparatively higher peak in the visible region with a wider absorption region than Br- and Cl-containing materials. Besides, with the substitution of Ge by Sn in the B-site, there is a drastic decrease in absorption peak as depicted in Figure 5. The extracted results for the Fr-based series are in line with the results found in literature for such ABX$_3$ lead-free perovskite materials.[24], [25], [28], [32]–[34], [36], [56] Generally, approximately 43% of the solar spectrum is covered by visible light, whereas only 4% of solar energy contribution comes from the ultraviolet light. Therefore, to take advantage of the visible light energy solar spectrum for photovoltaic conversion, intrinsic FrGeX$_3$ exhibits more suitability than FrSnX$_3$ from the application point of view.

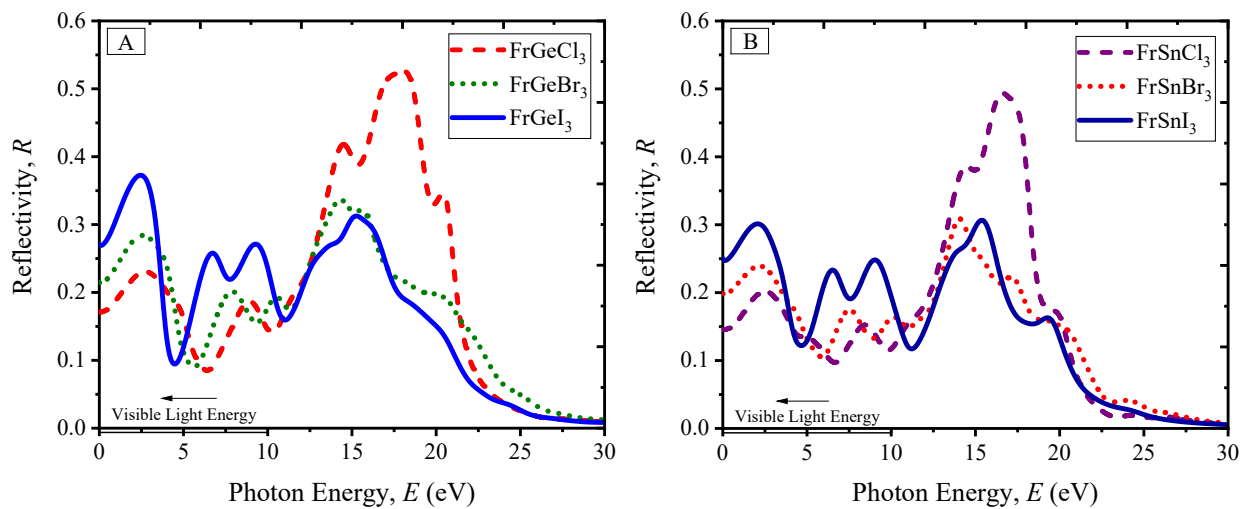

**Figure 6**. Reflectivity profile of perovskites (A) FrGeX$_3$ (X = I, Br, Cl), and (B) FrSnX$_3$ (X=I, Br, Cl)

Reflectivity is an optical property of materials to understand the surface nature of perovskites. The reflectivity determines how much light energy is reflected from the materials' surface compared to the light-energy incident on the surface. The reflectivity spectra of pristine FrBX$_3$ compounds are illustrated in Figure 6(A) for FrGeX$_3$ and 6(B) for FrSnX$_3$ compound, for photon energy up to 30eV. From Figure 6(A) and 6(B), it is observed that all the materials exhibit identical behavior with providing low reflectivity in the whole spectrum. Exceptionally, Cl- containing materials show an additional peak in the UV region, indicating the significant absorptivity and/or transmission. At the low energy region, FrBI$_3$ (B=Ge, Sn) materials reflect more energy compared to the other compound and the maximum reflectivity of 0.38 is observed for FrGeI$_3$. This moderate reflectivity at the visible region indicates the material to be a good candidate for solar cell applications. The dielectric variation of the investigated perovskite materials is illustrated in Figure. 7(A), for up to 30eV photon energy. The detected patterns indicate that the materials are highly transmittable in the high energy region as well as low transmittance in the low energy region. The dielectric constant values are used to determine how well optoelectronic devices work, defined as a material's response to incident light energy. The greater dielectric values at a lower charge carrier recombination rate are profound in improving the optoelectronic device performance in the long run. From both real and imaginary dielectrics, it is evident that materials comprising with I exhibit comparatively higher amplitude of dielectric constant than other samples in the visible region. Light propagation behavior in absorbing materials can be portrayed with the refractive index value. The greater the refractive index, the closer the light will travel to its normal direction. The refractive index of all studied perovskites is qualitatively similar, with a little variation in peak heights and placements as displayed in *supplementary Figure S9*. The static refractive index varies depending on the compound n(0), which is found a maximum of 3.15 for $FrGeI_3$, and the minimum value of 2.25 is found for $FrSnCl_3$. Besides, optically conductive materials also ascertain the photoconductivity, which increases electric conductivity with the rise of both photon absorption of electromagnetic radiation. The variations in optical conductivity for the pristine $FrBX_3$ compounds are shown in Figures 7(E) and 7(F). At low energies, it is clear that the majority of the compounds have a noticeable optical conductivity. The highest amplitudes are depicted for Cl-containing materials in the high-energy region, while I-containing compounds exhibit the opposite manner. However, Ge-containing materials exhibit higher conductivity compared to Sn-based samples, which confirms the better performance of $FrGeX_3$.

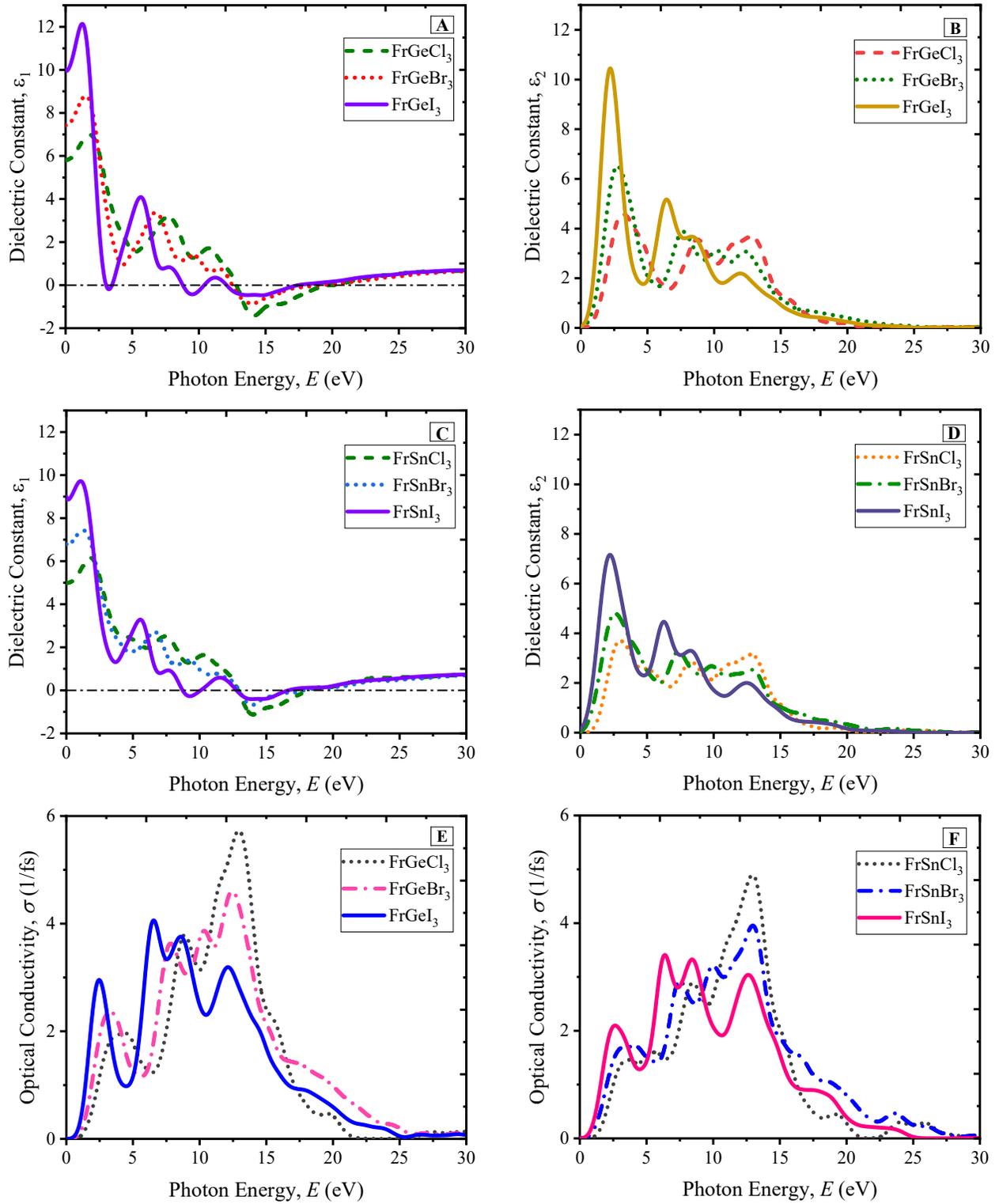

**Figure 7**. Real and imaginary part of the dielectric constant for FrGeX3 are presented in (A) and (B) respectively, and for FrSnX3, presented in (C) and (D). The optical conductivity of FrGeX3 with respect to the photon energy is presented in (E) and that of FrSnX3 in (F).

## 3.4 Mechanical properties

The elastic constants depict a clear understanding of the mechanical stability in any crystals. This study employed the CASTEP module to calculate the elastic constants using finite strain theory. The three independent elastic constants for the cubic perovskites $C_{11}$, $C_{12}$, and $C_{44}$ are evaluated using the CASTEP stress-strain method and listed in Table IV. As for the cubic symmetry, the Born-stability criterion as follows, $C_{11}>0$, $C_{44}>0$, $C_{11}-C_{12}>0$, and $C_{11}+2C_{12}>0$, should be met by perovskite materials. It is evident from Table III that these pristine $FrBX_3$ perovskite compounds agree well to bear this study's reliability. Observation of the Cauchy pressure ($C_{12}-C_{44}$) is an essential and valuable parameter for predicting the compounds' brittleness and ductility. From the retrieved values of Cauchy pressure, it is observed that all the pristine materials exhibiting positive values correspond to the ductile nature except for $FrGeI_3$ giving negative values. Hill approximations are used to estimate the most crucial mechanical parameters for the polycrystalline materials, such as bulk modulus B and shear modulus G using the single-crystal elastic constants. However, Young's modulus $Y$, Pugh's ratio B/G, and the Poisson's ratio $v$ of these pristine compounds are determined using the following formula,

$$Y = \frac{9BG}{(3B+G)} \text{ and } v = \frac{(3B-2G)}{(6B+2G)}.$$

Accordingly, bulk modulus (B) is one of the noteworthy mechanical parameters that provide an insight into the material's rigidity. All the investigated materials show flexibility and softness, as observed from the calculated bulk modulus values ranging between 14.50 GPa (for $FrSnI_3$) and 26.60 GPa (for $FrGeI_3$). Therefore, such metal-halide perovskite materials can be easily incorporated into thin films for solar cell applications due to their flexibility and softness. The change in halide ion from Cl to I results in a decrease in bulk modulus value for all the samples; where a change in B site with Ge from Sn leads to the opposite trend slightly, and a similar tendency is seen for shear modulus ($G$) and young modulus ($Y$). Pugh's ratio (B/G) and Poisson ratio $v$ are used to predict a material's failure mechanism, i.e., its ductility and brittleness. Pugh's ratio distinguishes ductile from brittle materials, and the Poisson ratio indicates ductile from brittle materials with a crucial value of 1.75 and 0.26, respectively. Pugh's ratio more than 1.75, or Poisson ratio greater than 0.26, indicates ductility; otherwise, it indicates brittleness. The criteria state that $FrGeI_3$ is brittle, $FrGeBr_3$ is at the brittle-ductile boundary, and the remaining compounds are ductile, as seen from the value in Table IV, while the Poisson ratio has the same characteristics. Hence, for the device applications, FrGeI3 would be challenging because its low Pugh's ratio and Poisson ratio indicate its' brittle nature. In contrast, compounds containing Cl have a higher degree of ductility, and $FrSnCl_3$ has the most ductility of the bunch.

*Table IV: The evaluated elastic constants $C_{ij}$ (GPa) and Cauchy Pressure $C_{12}$-$C_{44}$ (GPa) with the calculated mechanical properties of cubic $FrBX_3$ (B=Ge, Sn; X= Cl, Br, I) perovskites.*

| Phase | $C_{11}$ | $C_{12}$ | $C_{44}$ | $C_{12}$-$C_{44}$ | B (GPa) | G (GPa) | Y (GPa) | B/G | $v$ |
|---|---|---|---|---|---|---|---|---|---|
| $FrGeCl_3$ | 52.41 | 13.69 | 12.17 | 1.52 | 26.60 | 14.67 | 37.18 | 1.81 | 0.27 |
| $FrGeBr_3$ | 45.86 | 10.89 | 10.28 | 0.61 | 22.55 | 12.73 | 32.14 | 1.77 | 0.26 |
| $FrGeI_3$ | 33.17 | 6.55 | 7.51 | -0.96 | 15.42 | 9.46 | 23.56 | 1.63 | 0.24 |
| $FrSnCl_3$ | 48.41 | 9.82 | 6.54 | 3.28 | 22.68 | 10.27 | 26.77 | 2.21 | 0.30 |
| $FrSnBr_3$ | 42.11 | 7.72 | 5.37 | 2.35 | 19.18 | 8.75 | 22.78 | 2.19 | 0.30 |
| $FrSnI_3$ | 33.64 | 4.94 | 3.64 | 1.30 | 14.50 | 6.55 | 17.08 | 2.21 | 0.30 |

## *3.5 Environmentally-friendly (lead-free) perovskites*

Table V shows an overview of the essential features of the metal halide perovskites that are studied. Ge-based compounds, in comparison, show stronger optical absorbance and photoconductivity in the low-energy region than Sn-based compounds, implying a better B site element to be germanium (Ge) with Fr in A site. Furthermore, the optical absorbance and conductivity of the Sn-based compounds, $FrSnX_3$ (X=Cl, Br, I), are comparatively lower than those of Ge-based compounds. Henceforth, due to its strong absorption and photoconductivity throughout the solar spectrum, $FrGeI_3$ can be considered as the potential lead-free perovskite material for photovoltaic applications if one can take care of the small brittleness nature of this compound. Contrariwise, $FrGeCl_3$ is ductile but has a significant band gap, rendering it unsuitable for solar visible light absorption. However, by incorporating another halide ion in the X site and/or a metal in the B-site with FrGeI$_3$ following the solid solution or doping methods as well as other methods introduced in DFT literature, film formation may be improved by reducing the brittleness of the materials. [25], [29]–[34], [56]–[59] However, as a test calculation we have performed the solar cell device simulation using SCAPS 1D simulation software for the FrGeI3. In this simulation the absorption profile of the perovskite FrGeI3 is used which was extracted from the DFT simulation, and for HTL we have used CZTSe where TiO2 is used for ETL materials. As the contact materials, we used gold and FTO. The simulated I-V graph can be found in Fig. S10 in the supplementary material with other relevant parameters. With the extracted parameters, the PCE is found to be 15.07% which is subjected to further optimizations.

*Table V: Summary of key properties on the investigated perovskites*

| Properties | Optical absorption | Photoconductivity | Failure mode |
|---|---|---|---|
| FrGeCl$_3$ | High [UV-region] <br> Less [Visible region] | Moderate | Ductile |
| FrGeBr$_3$ | High [UV-region] <br> Medium [Visible-region] | Moderate | Ductile |
| FrGeI$_3$ | High [UV-region] <br> High [Visible-region] | High | Brittle |
| FrSnCl$_3$ | High [UV-region] <br> Less [Visible-region] | Less | Ductile |
| FrSnBr$_3$ | High [UV-region] <br> Medium [Visible-region] | Moderate | Ductile |
| FrSnI$_3$ | High [UV-region] <br> Medium [Visible-region] | Moderate | Ductile |

## Conclusion

FrBX$_3$ (B = Sn, Ge; X = I, Br, Cl), a Pb-free inorganic metal-halide (MH) cubic perovskite was studied using the first-principle of DFT simulations to evaluate its structural, electrical and optical properties. A decrease of electronic band gap (E$_g$) is observed while using Cl, Br, I in the halogen site, respectively, and that tends to be a slight change in other properties (variations in the halogen contents in these materials can be used to modulate the bandgap). Ge appears to be a superior candidate than Sn as Ge-based compounds exhibit higher optical absorption and optical conductivity compared to that Sn-based materials. Though Fr is radioactive, the studied materials are found to be mechanically stable in nature and can be grown into thin films easily due to their low bulk modulus. Considering the retrieved structural, mechanical and optoelectronic properties, it is inferred that FrBX$_3$ (B=Ge, Sn; X = Cl, Br, I) would be an excellent candidate as a metal-halide perovskite for various photovoltaic applications including photoconductor-based X-ray detector. According to the existing knowledge in the literature, a further study based on the doping of different cations in B- and X-sites along with the pressure-induced study may resolve the issue.

## Acknowledgment

The authors are grateful to the center of excellence of the Department of Mathematics and Physics, North South University (NSU), Dhaka 1229, Bangladesh. This research is funded by the NSU research grant CTRG-20/SEPS/13. The authors are grateful to Dr. Mozahar Ali and Mr. Sadiq Shahriyar Nishat for their useful suggestions regarding the DFT calculation.